\documentclass[10pt,showpacs,amsmath,amssymb]{revtex4}
       
\usepackage{graphicx}
\usepackage{bbold}
\usepackage{epsfig}
\usepackage{rotating}
\usepackage{graphpap}

\begin{document}
\setlength{\unitlength}{1mm}

\title{Self-Organized Formation of Retinotopic Projections Between Manifolds of Different 
Geometries --  
Part 3: Spherical Geometries}

\author{M. G{\"u}{\ss}mann}
\email{martin.guessmann@itp1.uni-stuttgart.de}
\affiliation{1.$\,$Institut f{\"u}r Theoretische Physik, Universit{\"a}t Stuttgart, 
Pfaffenwaldring 57, 70569 Stuttgart, Germany}
\author{A. Pelster}
\email{axel.pelster@uni-due.de}
\affiliation{Fachbereich Physik, Campus Duisburg, Universit{\"a}t Duisburg-Essen, 
Lotharstrasse 1, 47048 Duisburg, Germany}
\author{G. Wunner}
\email{guenter.wunner@itp1.uni-stuttgart.de}
\affiliation{1.$\,$Institut f{\"u}r Theoretische Physik, Universit{\"a}t Stuttgart, 
Pfaffenwaldring 57, 70569 Stuttgart, Germany}

\date{\today}

\begin{abstract}
We follow our general model in Ref.~\cite{gpw1} and analyze the 
formation of retinotopic projections for 
the biologically relevant situation of {\it spherical} geometries.
To this end we elaborate both a linear and a nonlinear synergetic analysis which results in 
order parameter equations for the dynamics of connection weights between two spherical 
cell sheets. We show that these equations of evolution provide stable stationary solutions 
which correspond to retinotopic modes. 
A further analysis of higher modes furnishes proof
that our model describes the emergence of a perfect one-to-one retinotopy between two spheres. 
\end{abstract}
\maketitle
\section{Introduction} \label{intro}
An essential precondition for a correct operation of the nervous system consists
in well-ordered neural connections between different cell sheets. An example, which has been
explored both experimentally and theoretically in detail, is the formation of ordered
projections between retina and tectum, a part of the brain which plays an 
important role in processing optical information \cite{goodhill}.
At an initial stage of ontogenesis, retinal ganglion cells have random synaptic 
contacts with the tectum. In the adult animal, however, a so-called {\it retinotopic} 
projection is realized: Neighboring cells of the retina project onto neighboring
cells of the tectum.
A detailed analytical treatment of H{\"a}ussler and von der Malsburg described
these ontogenetic processes in terms of self-organization \cite{Malsburg}. In that work
retina and tectum were treated as one-dimensional discrete cell arrays.
The dynamics of the connection weights between retina and tectum were assumed to be
governed by the so-called H{\"a}ussler equations. In Ref.~\cite{gpw1} we generalized 
these equations of evolution to {\it continuous} manifolds of {\it arbitrary geometry} 
and {\it dimension}. Furthermore, we performed an extensive synergetic analysis 
\cite{Haken1,Haken2} 
near the instability of stationary uniform connection weights between retina and tectum.
The resulting generic order parameter equations served as a starting point for 
analyzing retinotopic projections 
between Euclidean manifolds in Ref.~\cite{gpw2}. Our results for strings 
turned out to be analogous to those for discrete linear chains, i.e. our model included 
the special case of H{\"a}ussler and von der Malsburg \cite{Malsburg}. Additionally, 
we could show in the case of planar geometries that superimposing two modes under suitable 
conditions provides a state with a pronounced retinotopic character.\\ 

\begin{figure}[t!]
\centerline{\includegraphics[scale=0.75]{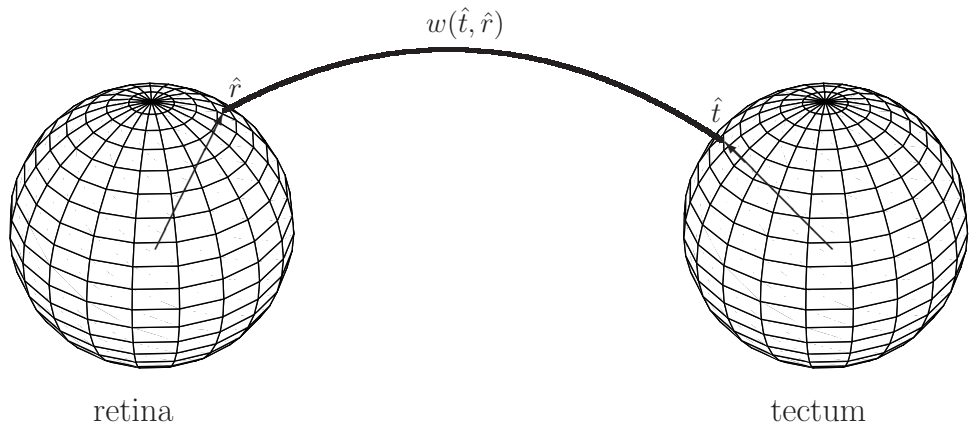}}
\caption[Kugel]{\label{kugel} \small The cells of retina and tectum, which are assumed to 
be continuously distributed on unit spheres, are represented by their unit vectors 
$\hat r$ and $\hat t$, respectively. 
The two cell sheets are connected by positive connection weights $w(\hat t,\hat r)$.}
\end{figure}

In this paper we apply our general model \cite{gpw1} again to projections between two-dimensional
manifolds. Now, however, we consider manifolds with {\it constant positive curvature}. 
Typically, the retina represents approximately a hemisphere, whereas the tectum has 
an oval form \cite{goodhill}. 
Thus, it is biologically reasonable to model both cell sheets by spherical manifolds. 
Without loss of generality we assume that the two cell sheets for retina and tectum
are represented by the surfaces of two unit spheres, respectively. Thus, in our model, 
the corresponding continuously distributed cells are represented by unit vectors 
$\hat r$ and $\hat t$. Every ordered pair $(\hat t,\hat r)$ 
is connected by a positive connection weight $w(\hat t,\hat r)$ 
as is illustrated in Figure~\ref{kugel}. 
The generalized H{\"a}ussler equations of Ref.~\cite{gpw1,thesis} for these connection 
weights are specified as follows
\begin{equation} 
\label{hslerkugel}
\dot w(\hat t,\hat r)=f(\hat t,\hat r,w)-\frac{w(\hat t,\hat r)}{8\pi} 
\hspace*{2mm} \int \! d\Omega_{t'}\,f(\hat t\,',\hat r,w)
-\frac{w(\hat t,\hat r)}{8\pi}\hspace*{2mm} 
\int \! d\Omega_{r'}\,f(\hat t,\hat r\,',w)\,.
\end{equation}
The first term on the right-hand side describes cooperative synaptic growth processes, 
and the other terms stand for corresponding competitive growth processes. The total 
growth rate is defined by
\begin{equation}
\label{GRO}
f(\hat t,\hat r,w)=\alpha+w(\hat t,\hat r) \int \! d\Omega_{t'}
\int \! d\Omega_{r'} c_T(\hat t \cdot \hat t\,')\,
c_R(\hat r \cdot \hat r\,')\,w(\hat t\,',\hat r\,')\,,
\end{equation}
where $\alpha$ denotes the global growth rate of new synapses onto the tectum, and 
is the control parameter of our system. The cooperativity functions $c_T(\hat t \cdot \hat t\,')$, 
$c_R(\hat r \cdot \hat r\,')$ represent the neural connectivity within each manifold. 
They are assumed to be 
positive, symmetric with respect to their arguments, and normalized. 
The integrations in (\ref{hslerkugel}) and (\ref{GRO}) are performed over all
points $\hat t, \hat r$ on the manifolds, where 
$d\Omega_t,d\Omega_r$ represent the differential solid angles of the
corresponding unit spheres. 
Note that the factors $8\pi$ in Eq.~(\ref{hslerkugel}) are twice the measure $M$ of the 
unit sphere, which is given by
\begin{equation} \label{kumass}
M=\int\!\! d\Omega_t=\int\!\! d\Omega_r=\int\limits_0^{2\pi}\!\!d\varphi\int\limits_0^{\pi}\!\!\sin \vartheta d\vartheta=4\pi\,.
\end{equation}
If the global growth rate of new synapses onto the tectum $\alpha$ is large enough, the 
long-time dynamics
is determined by a uniform connection weight. However, we shall see within a linear
analysis in Section~\ref{linanalys} that 
this stationary solution becomes unstable at a critical value of the global
growth rate.
Therefore, we have to perform a nonlinear synergetic analysis, in Section~\ref{nonlinanalys}, 
which yields the underlying order parameter equations
in the vicinity of this bifurcation. As in the case of Euclidean manifolds, 
we show that they have no quadratic terms, represent a potential dynamics,
and allow for retinotopic modes.
In Section~\ref{11retino} we include the influence of higher modes upon the connection weights, 
which leads to
recursion relations for the corresponding amplitudes.
If we restrict ourselves to special cooperativity functions, 
the resulting recursion relations can be solved analytically by using the method of 
generating functions. As a result of our analysis we obtain a perfect one-to-one
retinotopy if the global growth rate $\alpha$ is decreased to zero.

\section{Linear Analysis} \label{linanalys}
According to the general reasoning in Ref.~\cite{gpw1} we start with fixing the metric 
on the manifolds and determine the eigenfunctions
of the corresponding Laplace-Beltrami operator. Afterwards, we expand the 
cooperativity functions with respect to these eigenfunctions
and perform 
a linear analysis of the stationary uniform state.  

\subsection{Laplace-Beltrami Operator}
For the time being we neglect the distinction between retina and tectum,
because the following considerations are valid for both manifolds. 
Using spherical coordinates, we write
the unit vector on the sphere as $\hat x=(\sin\vartheta \cos\varphi,\sin\vartheta \sin\varphi,\cos\vartheta)$.
The Laplace-Beltrami operator on a manifold reads quite generally \cite{klein}
\begin{equation}
\label{LBO}
\Delta=\frac{1}{\sqrt{g}} \, \partial_{\lambda} \left( g^{\lambda\mu} \sqrt{g} \,\partial_{\mu}\right)\,.
\end{equation}
For the sphere the components of the covariant tensor $g_{\mu\nu}$ are
\begin{equation} \label{kulinel}
g_{11}=\left(\frac{\partial\hat x}{\partial \vartheta}\right)^2=1\,,\quad 
g_{12}=g_{21}=\frac{\partial \hat x}{\partial \vartheta}\,\frac{\partial \hat x}{\partial \varphi}=0\,,\quad g_{22}=
\left(\frac{\partial \hat x}{\partial \varphi}\right)^2=\sin^2\vartheta\,.
\end{equation}
With this the determinant of the covariant metric tensor reads $g=\sin^2\vartheta$ and the components 
of the contravariant metric are given by
\begin{equation}
g^{11}=1\,,\quad g^{12}=g^{21}=0\,,\quad g^{22}=\frac{1}{\sin^2\vartheta}\,,
\end{equation}
whence the Laplace-Beltrami operator 
for the sphere takes the well-known form
\begin{equation} \label{einbettwinkel}
\Delta_{\vartheta,\varphi}=\frac{1}{\sin\vartheta}\,\frac{\partial}{\partial\vartheta}\left(\sin\vartheta \frac{\partial}{\partial\vartheta}\right)+
\frac{1}{\sin^2\vartheta}\frac{\partial^2}{\partial\varphi^2}\,.
\end{equation}
Its eigenfunctions are
known to be given by spherical harmonics $Y_{lm}(\hat x)$:
\begin{equation}
\Delta_{\vartheta,\varphi}\,Y_{lm}(\hat x)=-l(l+1)Y_{lm}(\hat x)\,.
\end{equation}
With $l=0,1,2,\ldots$ and $m=-l,-l+1,\ldots,l-1,l$ they are $(2l+1)$-fold degenerate and 
form a complete orthonormal system on the unit sphere:
\begin{eqnarray}
\int\!\! d\Omega_x\,Y_{lm}(\hat x\,)Y_{l'm'}^*(\hat x\,)&=&\delta_{l^{ }l'}\delta_{m^{ }m'}\,, \label{KUOR1}\\
\sum_{l=0}^{\infty}\sum_{m=-l}^l Y_{lm}(\hat x\,)Y_{lm}^*(\hat x'\,)&=&\delta(\hat x-\hat x')\,.
\end{eqnarray}

\subsection{Cooperativity Functions}

The argument of the cooperativity functions $c(\hat x \cdot \hat x')$ is
the scalar product $\hat x\cdot \hat x'$ which takes values between $-1$ and $+1$. Therefore the 
cooperativity functions can be expanded in terms of
Legendre functions $P_l(\hat x\cdot \hat x')$, which form a complete orthogonal system on this 
interval \cite[7.221.1]{grad}: 
\begin{eqnarray}
\int\limits_{-1}^{1}P_l(\sigma)P_{l'}(\sigma)\,d\sigma&=&\frac{2}{2l+1}\,\delta_{ll'}\,,\label{wagnerportho}\\
\frac{1}{2}\sum_{l=0}^{\infty}(2l+1)P_l(\sigma')P_l(\sigma)&=&\delta(\sigma-\sigma')\,.\label{wagnerpvollst}
\end{eqnarray}
Then the expansion of the cooperativity functions read
\begin{equation}
c(\hat x \cdot \hat x')=\sum_{l=0}^{\infty}\frac{2l+1}{4\pi}f_l\,P_l(\hat x \cdot \hat x')\,,\label{kuct}
\end{equation}
where $f_l$ denote the respective expansion coefficients. Using the Legendre addition theorem 
\cite{arf}
\begin{equation} \label{kuaddtheorem}
P_l(\hat x \cdot \hat x')=\frac{4\pi}{2l+1}\sum_{m=-l}^l Y_{lm}(\hat x\,)Y_{lm}^*(\hat x')\,,
\end{equation}
we arrive, for each manifold, at the expansion
\begin{equation} \label{KU1}
c_T(\hat t \cdot \hat t')=\sum_{L=0}^{\infty}\sum_{M=-L}^L f_L^T Y_{LM}^T(\hat t\,) Y_{LM}^{T*}(\hat t'\,)\,,
\quad c_R(\hat r \cdot \hat r')=\sum_{l=0}^{\infty}\sum_{m=-l}^l f_l^R Y_{lm}^R(\hat r) Y_{lm}^{R*}(\hat r'\,)\,.
\end{equation}
Note that the normalization of the cooperativity functions and the orthonormality relations
(\ref{KUOR1}) lead to the constraints $f_0^T=f_0^R=1$. 

\subsection{Eigenvalues}

The initial state of ontogenesis with randomly distributed synaptic contacts is 
described by the stationary uniform solution of the generalized H{\"a}ussler equations, 
$w_0(\hat t,\hat r)=1$. Its stability is analyzed by linearizing the H{\"a}ussler equations 
(\ref{hslerkugel})
with respect to the deviation $v(\hat t,\hat r)=w(\hat t,\hat r)-w_0(\hat t,\hat r)$.
The resulting linearized equations read
\begin{equation} 
\label{gen43}
\dot v(\hat t,\hat r)=\hat{L}(\hat t,\hat r,v) 
\end{equation}
with the linear operator 
\begin{eqnarray}
\label{Loperator}
&&\hspace*{0.5cm}\hat{L}(\hat t,\hat r,v)=-\alpha v (\hat t,\hat r)
-\frac{1}{8\pi}\int\! d\Omega_{t'} \left[v(\hat t',\hat r) + 
\int\! d\Omega_{t''}\int\! d\Omega_{r''}\, c_T(\hat t'\cdot\hat t'') \, c_R(\hat r\cdot\hat r'')\, v(\hat t'',\hat r'')
\right]
\nonumber \\
&&\hspace{-0.8cm}-\frac{1}{8\pi}\int\! d\Omega_{r'} \left[v(\hat t,\hat r') + 
\int\! d\Omega_{t''}\!\int\! d\Omega_{r''}\, c_T(\hat t\cdot\hat t'') \, c_R(\hat r' \cdot\hat r'')\, v(\hat t'',\hat r'')
\right]+\int\! d\Omega_{t'}\!\int\! d\Omega_{r'}\, c_T(\hat t\cdot \hat t') \, c_R(\hat r\cdot \hat r')\, 
v(\hat t',\hat r')
\,.
\end{eqnarray}
To solve Eq.~(\ref{gen43}), we have to consider the eigenvalue problem of 
the linear operator (\ref{Loperator}). It has the eigenfunctions
\begin{equation} \label{efsphere}
v_{L l}^{M m}(\hat t,\hat r)=Y_{LM}^T(\hat t\,) Y_{lm}^R(\hat r)
\end{equation}
and the spectrum of eigenvalues reads \cite{gpw1}:
\begin{equation} \label{kuew1042}
\Lambda_{Ll}^{Mm}=\left\{\begin{array}{cc}
-\alpha-1&L=M=l=m=0\\
-\alpha+(f_L^T f_l^R-1)/2&L=M=0,\,(l,m)\not=(0,0)\\
 &l=m=0,\,(L,M)\not=(0,0)\\
-\alpha+f_L^T f_l^R& \mbox{ otherwise }.
\end{array} \right.
\end{equation}
By changing the uniform growth rate $\alpha$ in a suitable way, 
the real parts of some eigenvalues (\ref{kuew1042}) become positive
and the system can be driven to the neighborhood of an instability. 
Which eigenvalues (\ref{kuew1042}) become unstable
in general depends on the respective values
of the given expansion coefficients $f_L^T$, $f_l^R$. 
If we assume monotonically decreasing expansion coefficients $f_L^T$, $f_l^R$,
\begin{equation}
1=f_0^T\geq f_1^T\geq f_2^T\geq \cdots \geq 0\,,\qquad 1=f_0^R\geq f_1^R\geq f_2^R\geq \cdots \geq 0\,,
\end{equation}
the maximum eigenvalue in (\ref{kuew1042}) is given by
$\Lambda_{\rm max}=\Lambda_{11}^{M m}=-\alpha+f_1^T f_1^R$.
Thus, the instability occurs when the global growth rate reaches
its critical value $\alpha_c=f_1^T f_1^R$.
At this instability point all nine modes with $(L^u,l^u)=(1,1)$ and $M^u=0,\pm 1$, 
$m^u=0,\pm 1$ become unstable, where we have introduced the index $u$ for the unstable modes. 

\section{Nonlinear Analysis} \label{nonlinanalys}
In this section we specialize the generic order parameter equations of Ref.~\cite{gpw1}
to unit spheres. We observe
that the quadratic term vanishes and derive selection rules for the appearance of cubic
terms. Furthermore, we essentially simplify the calculation of the order parameter 
equations by taking into account the symmetry properties of the cubic terms.
We show that the order parameter equations
represent a potential dynamics, and determine the underlying potential. 
\subsection{General Structure of Order Parameter Equations}
The linear stability analysis motivates treating the nonlinear H{\"a}ussler equations 
(\ref{hslerkugel})
near the instability by decomposing the deviation $v(\hat t,\hat r)=w(\hat t,\hat r)-
w_0(\hat t,\hat r)$
in unstable and stable contributions, 
\begin{equation}
\label{US}
v(\hat t,\hat r) = U(\hat t,\hat r)+S(\hat t,\hat r)\,.
\end{equation}
Using Einstein's sum convention the expansion of the unstable modes reads
\begin{equation}
\label{UEXP}
U(\hat t,\hat r)=
U_{11}^{M^u m^u} Y_{1 M^u}^T (\hat t\,) Y_{1 m^u}^R (\hat r\,) \, ,
\end{equation}
and, correspondingly, the contribution of the stable modes is given by
\begin{equation}
\label{SEXP}
S(\hat t,\hat r)=
S_{L l}^{M m} Y_{L M}^T (\hat t\,) Y_{l m}^R (\hat r\,) \,.
\end{equation}
Note that the summation in (\ref{SEXP}) is performed over all parameters $(L,l)$ except 
for $(L^u,l^u)=(1,1)$, i.e. from now on the parameters $(L,l)$ stand for the stable modes alone. 
With the help of the 
slaving principle of synergetics \cite{Haken1,Haken2} the original high-dimensional 
system can be reduced 
to a low-dimensional one which only contains the unstable amplitudes. The 
resulting order parameter equations read \cite{gpw1}
\begin{eqnarray}
\hspace{-0.6cm}\dot U^{M^u m^u} &=&\Lambda \,
U^{M^u m^u} +
A_{M^u, M^u{}' M^u{}''}^{m^u, m^u{}' m^u{}''}
\,U^{M^u{}' m^u{}'} \, U^{M^u{}'' m^u{}''} 
+ B_{M^u, M^u{}' M^u{}'' M^u{}'''}^{m^u{},
m^u{}' m^u{}'' m^u{}'''} U^{M^u{}' m^u{}'}
\, U^{M^u{}'' m^u{}''}\, U^{M^u{}''' m^u{}'''} \,.
\label{OPE}
\end{eqnarray}
They contain, as usual, a linear, a quadratic, and a cubic term of the order parameters. 
The corresponding coefficients can be expressed in terms
of the expansion coefficients $f_L^T$, $f_l^R$
of the cooperativity functions (\ref{KU1}) and integrals over products of the eigenfunctions
$Y_{l m}(\hat x)$:
\begin{eqnarray} 
I_{l,l^{(1)} l^{(2)} \ldots  l^{(n)}}^{m,m^{(1)} m^{(2)} \dots m^{(n)}}
&=&\hspace*{1mm}\int\! d\Omega_x\,
Y_{lm}^*(\hat x)\,Y_{l^{(1)}\,m^{(1)}}(\hat x)\,Y_{l^{(2)}\,m^{(2)}}(\hat x)\,\cdots\,
Y_{l^{(n)}\,m^{(n)}}(\hat x)\,,\label{abkurzI}\\
J_{l^{(1)} l^{(2)} \ldots  l^{(n)}}^{m^{(1)} m^{(2)} \ldots m^{(n)}}
&=&\hspace*{1mm}\int\!d\Omega_x\,
Y_{l^{(1)}\,m^{(1)}}(\hat x)\, Y_{l^{(2)}\,m^{(2)}}(\hat x)\,\cdots\,
Y_{l^{(n)}\,m^{(n)}} (\hat x)\,. \label{abkurzJ}
\end{eqnarray}
The quadratic coefficients read
\begin{eqnarray}
\label{AA}
A_{M^u, M^u{}' M^u{}''}^{m^u, m^u{}' m^u{}''}  = f_1^T \,
f_1^R \, I_{1,1\,1}^{M^u,M^u{}' M^u{}''}
\, I_{1,1\,1}^{m^u,m^u{}' m^u{}''} \,,
\end{eqnarray}
whereas the cubic coefficients are
\begin{eqnarray}
B_{M^u, M^u{}' M^u{}'' M^u{}'''}^{m^u{}, m^u{}' m^u{}'' m^u{}'''}
&=& - \frac{1}{8\pi}\,f_1^T \, f_1^R
\left(I_{1,1\,1\,1}^{M^u,M^u{}' M^u{}'' M^u{}'''}
\,\delta_{m^u m^u{}'} \,J_{1\,1}^{m^u{}'' m^u{}'''}
+ I_{1,1\,1\,1}^{m^u,m^u{}' m^u{}'' m^u{}'''} \delta_{M^u M^u{}'} \,
J_{1\,1}^{M^u{}'' M^u{}'''} \right)
\nonumber \\&&
+\left\{ \left[ f_L^T \, f_l^R +f_1^T \, f_1^R \, \right]
I_{1,1\,L}^{M^u,M^u{}' M} \,
I_{1,1\,l}^{m^u,m^u{}' m}
- \frac{1}{4\sqrt{\pi}} \,
\left[\delta_{L0}\,\delta_{M0} \delta_{M^u M^u{}'}  \,
\left( 1+ f_l^R \right)
I_{1,1\,l}^{m^u,m^u{}' m}
\right.\right. \nonumber \\ && \left. \left.
+\delta_{l0}\,\delta_{m0}\delta_{m^u m^u{}'}
\, \left( 1+f_L^T\right)
\, I_{1,1\,L}^{M^u,M^u{}' M}
\right] \right\} H_{Ll}^{Mm,M^u{}'' m^u{}'' M^u{}''' m^u{}'''} \,
\,.
\label{BB}
\end{eqnarray}
Note that Eq.~(\ref{BB}) involves a summation over all stable modes $(L,M;l,m)$. As is 
common in synergetics, the cubic coefficients (\ref{BB}) consist in general 
of two parts, one stemming from the order parameters themselves and the other 
representing the influence of the center manifold $H$ on the order parameter dynamics 
according to
\begin{eqnarray}
\label{HN}
S_{L l}^{M m} = 
H_{Ll}^{Mm,M^u m^u M^u{}' m^u{}'} \,
U^{M^u m^u} U^{M^u{}' m^u{}'} \, .
\end{eqnarray}
Here the center manifold coefficients 
$H_{Ll}^{Mm,M^u m^u M^u{}' m^u{}'}$
are defined by 
\begin{eqnarray}
\label{HNN}
H_{Ll}^{Mm,M^u m^u M^u{}' m^u{}'}  &=& 
\frac{f_1^T f_1^R}{2\Lambda-\Lambda_{Ll} }
\Bigg[ I_{L,1\,1}^{M,M^u M^u{}'} \, 
I_{l,1\,1}^{m,m^u m^u{}'} 
- \frac{1}{4\sqrt{\pi}} \, \left(J_{1\,1}^{M^u M^u{}'} 
\,I_{l,1\,1}^{m,m^u{}' m^u{}''} \,\delta_{L0} \right.\nonumber\\  
 & &\left.+J_{1\,1}^{m^u m^u{}'} \,I_{L,1\,1}^{M,M^u M^u{}'} 
\,\delta_{l0} \right) \Bigg] \,. 
\label{QRES}
\end{eqnarray}

\subsection{Integrals}
The order parameter equations contain the following integrals: $J_{11}^{m' m''},
I_{1,11}^{m,m' m''},I_{l,11}^{m,m' m''},I_{1,1 l}^{m,m' m''},I_{1,111}^{m,m' m'' m'''}$.
The first integral is obtained by the orthonormality relation
(\ref{KUOR1}) and
\begin{equation} \label{kuyminusm}
Y_{l-m}(\hat x)=(-1)^m Y_{lm}^*(\hat x)\,,
\end{equation}
yielding $J_{1 1}^{m' m''}=(-1)^{m'}\delta_{m',-m''}$. Integrals over three and four spherical 
harmonics can be calculated with the help of the following relation \cite{cohen}:
\begin{equation}
Y_{l_1,m_1}(\hat x)Y_{l_2,m_2}(\hat x)=\sum_{l_3=|l_1-l_2|}^{l_1+l_2}
\sum_{m_3=-l_3}^{l_3}\sqrt{\frac{(2l_1+1)(2l_2+1)}{4\pi(2l_3+1)}}\,C(l_1,0,l_2,0|l_3,0)\,C(l_1,m_1,l_2,m_2|l_3,m_3)\,Y_{l_3,m_3}(\hat x)\,,\label{kuYprod}
\end{equation} 
where $C(l_1,m_1,l_2,m_2|l_3,m_3)$ represent the
Clebsch-Gordan coefficients \cite{heine}. 
Applying (\ref{kuYprod}) to integrals over three spherical harmonics
leads to
\begin{equation} \label{3fachint}
I_{l,l' l''}^{m,m' m''}=\sqrt{\frac{(2l'+1)(2l''+1)}{4\pi(2l+1)}}\,C(l',0,l'',0|l,0)\,
C(l',m',l'',m''|l,m)\,.
\end{equation}
For $l'=l''=1$ it follows
\begin{equation} \label{1046}
I_{l,1 1}^{m,m' m''}=\frac{3}{\sqrt{4\pi(2l+1)}}\,C(1,0,1,0|l,0)\,C(1,m',1,m''|l,m)\,.
\end{equation}
As the Clebsch-Gordan coefficients $C(l_1,0,l_2,0|l_3,0)$ vanish if the sum
$l_1+l_2+l_3$ is odd \cite{heine},
we obtain $I_{1,11}^{m,m' m''}=0$.
Thus, the quadratic contribution (\ref{AA}) to the
order parameter equations (\ref{OPE}) vanishes, by analogy with Euclidean manifolds \cite{gpw2}.
Furthermore, non-vanishing integrals (\ref{1046}) can only occur for $l=0$ and $l=2$. 
For $l=0$ we obtain
from the Clebsch-Gordan coefficients \cite{heine}
the result
\begin{equation}
I_{0,1 1}^{0,m' m''}=\frac{(-1)^{m'}}{\sqrt{4\pi}}\,\delta_{m',m''}\,.
\end{equation}
For $l=2$ we find, correspondingly, the nonvanishing integrals
\[
I_{0,11}^{0,00}=\frac{1}{\sqrt{4\pi}}\,,\quad I_{2,11}^{0,00}=\frac{1}{\sqrt{5\pi}}\,,
\quad I_{0,11}^{0,1-1}=I_{0,11}^{0,-11}=-\frac{1}{\sqrt{4\pi}}\,,\quad I_{2,11}^{0,1-1}=
I_{2,11}^{0,-11}=\frac{1}{\sqrt{20\pi}}\,,
\]
\begin{equation}
I_{2,11}^{1,10}=I_{2,11}^{1,01}=I_{2,11}^{-1,-10}=I_{2,11}^{-1,0-1}=\frac{3}{2\sqrt{15\pi}}
\,,\quad I_{2,11}^{2,11}=\frac{3}{\sqrt{30\pi}}\,,\quad I_{2,11}^{-2,-1 -1}=-\frac{3}{\sqrt{30\pi}}\,.
\end{equation}
Furthermore, the integrals $I_{1,1 l}^{m,m' m''}$ follow from
\begin{equation} \label{ku1064}
I_{1,1 l}^{m,m' m''}=(-1)^{m'+m''}I_{l,1 1}^{-m'',-m\,m'}\,.
\end{equation}
Integrals over four spherical harmonics can also be calculated with the help of 
(\ref{kuYprod}), and the result is
\begin{equation} \label{sphere38}
I_{l,l' l'' l'''}^{m,m' m'' m'''}=\sum_{l_3=|l''-l'''|}^{l''+l'''}\sum_{m_3=-l_3}^{l_3}
\sqrt{\frac{(2l''+1)(2l'''+1)}{4\pi(2l_3+1)}}\,C(l'',0,l''',0|l_3,0)
C(l'',m'',l''',m'''|l_3,m_3) I_{l,l' l_3}^{m,m' m_3}\,.
\end{equation} 
Specialyzing (\ref{sphere38}) to $l=l'=l''=l'''=1$ and taking into account (\ref{3fachint}) 
leads to
$I_{1,1\,1\,1}^{m,m' m'' m'''}\propto\delta_{m'+m''+m''',m}$.
Thus, we obtain the selection rule that the nonvanishing integrals 
$I_{1,1\,1\,1}^{m,m' m'' m'''}$ fulfill the condition $m'+m''+m'''=m$. The detailed evaluation
yields for those the respective values
\begin{eqnarray}
I_{1,111}^{0,000}&=&\frac{9}{20\pi}\,,\nonumber\\
\hspace{-0.5cm}I_{1,111}^{0,1-10}=I_{1,111}^{0,-110}=I_{1,111}^{0,10-1}=I_{1,111}^{0,-101}
=I_{1,111}^{0,01-1}=I_{1,111}^{0,0-11}&=&-\frac{3}{20\pi}\,,\nonumber\\
\hspace{-0.5cm}I_{1,111}^{1,100}=I_{1,111}^{1,010}=I_{1,111}^{1,001}=I_{1,111}^{-1,-100}
=I_{1,111}^{-1,0-10}=I_{1,111}^{-1,00-1}&=&\frac{3}{20\pi}\,,\nonumber\\
I_{1,111}^{1,11-1}=I_{1,111}^{1,1-11}=I_{1,111}^{1,-111}=I_{1,111}^{-1,1-1-1}
=I_{1,111}^{-1,-11-1}=I_{1,111}^{-1,-1-11}&=&-\frac{3}{10\pi}\,.\label{kui4b}
\end{eqnarray}

\subsection{Order Parameter Equations} 
To simplify the calculation of the cubic coefficients (\ref{BB}) in the order parameter 
equations (\ref{OPE}), we perform some 
basic considerations which lead to helpful symmetry properties.
To this end we start with replacing $m^u$ by $-m^u$. Using Eq.~(\ref{kuyminusm}) we obtain 
$I_{1,1\,1\,1}^{m^u,m^{u'} m^{u''} m^{u'''}}=I_{1,1\,1\,1}^{-m^u,-m^{u'}-m^{u''}-m^{u'''}}$. 
Corresponding symmetry relations can also be derived for the other
terms in (\ref{BB}). Therefore, we conclude that the order parameter equation for 
$U^{-M^u -m^u}$ is
obtained from that of $U^{M^u m^u}$ by negating all 
indices $M^u$ and $m^u$ with unchanged factors. 
Thus, instead of explicitly calculating nine order parameter equations, 
it is sufficient to restrict oneself determining the order parameter equations for 
$U^{00}$, $U^{10}$, $U^{01}$, and $U^{11}$. The remaining five order parameter equations 
follow instantaneously from those
by applying the symmetry relations.  
With this the order parameter equations result in
\begin{eqnarray}
\dot U^{00}&=&\Lambda U^{00}+\beta_1(U^{00})^3-2\beta_2 U^{00}U^{-10}U^{10}-2\bar\beta_2 
U^{00}U^{0-1}U^{01}+2\beta_3 U^{00}U^{1-1}U^{-11}+2\beta_3 U^{00}U^{-1-1}U^{11}\nonumber\\
 & &+\beta_4U^{01}U^{10}U^{-1-1}+\beta_4U^{0-1}U^{-10}U^{11}+\beta_4U^{0-1}U^{10}U^{-11}+
\beta_4U^{01}U^{-10}U^{1-1}\,,\nonumber\\
\dot U^{11}&=&\Lambda U^{11}+\beta_4 U^{00}U^{01}U^{10}+\beta_5(U^{01})^2U^{1-1}+\beta_6U^{01}U^{0-1}U^{11}
+\beta_3(U^{00})^2U^{11}\nonumber\\
 & &+\beta_5(U^{10})^2U^{-11}+\bar \beta_6 U^{10}U^{-10}U^{11}+\beta_7U^{11}U^{1-1}U^{-11}+\beta_8(U^{11})^2
U^{-1-1}\,,\nonumber\\
\dot U^{-1-1}&=&\Lambda U^{-1-1}+\beta_4 U^{00}U^{0-1}U^{-10}+\beta_5(U^{0-1})^2U^{-11}+\beta_6U^{0-1}U^{01}
U^{-1-1}+\beta_3(U^{00})^2U^{-1-1}\nonumber\\
 & &+\beta_5(U^{-10})^2U^{1-1}+\bar \beta_6 U^{-10}U^{10}U^{-1-1}+\beta_7U^{-1-1}U^{-11}U^{1-1}
+\beta_8(U^{-1-1})^2U^{11}\,,\nonumber\\
\dot U^{1-1}&=&\Lambda U^{1-1}+\beta_4 U^{00}U^{0-1}U^{10}+\beta_5(U^{0-1})^2U^{11}+\beta_6U^{0-1}U^{01}U^{1-1}
+\beta_3(U^{00})^2U^{1-1}\nonumber\\
 & &+\beta_5(U^{10})^2U^{-1-1}+\bar \beta_6 U^{10}U^{-10}U^{1-1}+\beta_7U^{1-1}U^{11}U^{-1-1}
+\beta_8(U^{1-1})^2U^{-11}\,,\nonumber\\
\dot U^{-11}&=&\Lambda U^{-11}+\beta_4 U^{00}U^{01}U^{-10}+\beta_5(U^{01})^2U^{-1-1}+\beta_6U^{01}U^{0-1}U^{-11}
+\beta_3(U^{00})^2U^{-11}\nonumber\\
 & &+\beta_5(U^{-10})^2U^{11}+\bar \beta_6 U^{-10}U^{10}U^{-11}+\beta_7U^{-11}U^{-1-1}U^{11}
+\beta_8(U^{-11})^2U^{1-1}\,,\nonumber\\
\dot U^{01}&=&\Lambda U^{01}+\bar\beta_2U^{01}(U^{00})^2+\beta_9(U^{01})^2U^{0-1}-2\beta_3 U^{01}U^{10}U^{-10}
-\beta_4U^{00}U^{11}U^{-10}\nonumber\\
 & &-\beta_4U^{00}U^{10}U^{-11}-\beta_6 U^{01}U^{11}U^{-1-1}-\beta_6U^{01}U^{1-1}U^{-11}
-2\beta_5 U^{0-1}U^{11}U^{-11}\,,\nonumber\\
\dot U^{0-1}&=&\Lambda U^{0-1}+\bar\beta_2U^{0-1}(U^{00})^2+\beta_9(U^{0-1})^2U^{01}-2\beta_3 U^{0-1}U^{-10}
U^{10}-\beta_4U^{00}U^{-1-1}U^{10}\nonumber\\
 & &-\beta_4U^{00}U^{-10}U^{1-1}-\beta_6U^{0-1}U^{-1-1}U^{11}-\beta_6U^{0-1}U^{-11}U^{1-1}-2\beta_5 U^{01}
U^{-1-1}U^{1-1}\,,\nonumber\\
\dot U^{10}&=&\Lambda U^{10}+\beta_2U^{10}(U^{00})^2+\bar \beta_9(U^{10})^2U^{-10}-2\beta_3 U^{10}U^{01}U^{0-1}
-\beta_4U^{00}U^{11}U^{0-1}\nonumber\\
 & &-\beta_4U^{00}U^{01}U^{1-1}-\bar \beta_6U^{10}U^{11}U^{-1-1}-\bar \beta_6U^{10}U^{-11}U^{1-1}-2\beta_5 
U^{-10}U^{11}U^{1-1}\,,\nonumber\\
\dot U^{-10}&=&\Lambda U^{-10}+\beta_2U^{-10}(U^{00})^2+\bar \beta_9(U^{-10})^2U^{10}-2\beta_3 U^{-10}U^{0-1}
U^{01}-\beta_4U^{00}U^{-1-1}U^{01}\nonumber\\
 & &-\beta_4U^{00}U^{0-1}U^{-11}-\bar \beta_6U^{-10}U^{-1-1}U^{11}-\bar \beta_6U^{-10}U^{1-1}U^{-11}-
2\beta_5 U^{10}U^{-1-1}U^{-11}\,.\label{Uequ}
\end{eqnarray}
With the abbreviations $\tilde\gamma=\gamma/\pi^2$, $\gamma=f_1^T f_1^R$ and $\gamma^{L,l}=f_L^T f_l^R$
the respective coefficients in (\ref{Uequ}) read
\begin{eqnarray}
\beta_1&=&-\frac{9}{80}\tilde\gamma
+\frac{\tilde\gamma}{80}\frac{2\gamma+\gamma^{2,0}-1}{2\gamma-\alpha-(\gamma^{2,0}-1)/2}
+\frac{\tilde\gamma}{80}\frac{2\gamma+\gamma^{0,2}-1}{2\gamma-\alpha-(\gamma^{0,2}-1)/2}
+\frac{\tilde\gamma}{25}\frac{\gamma+\gamma^{2,2}}{2\gamma-\alpha-\gamma^{2,2}}\,,
\nonumber\\
\beta_2&=&-\frac{9}{80}\tilde\gamma
-\frac{\tilde\gamma}{40}\frac{2\gamma+\gamma^{2,0}-1}{2\gamma-\alpha-(\gamma^{2,0}-1)/2}
+\frac{\tilde\gamma}{80}\frac{2\gamma+\gamma^{0,2}-1}{2\gamma-\alpha-(\gamma^{0,2}-1)/2}
-\frac{2\tilde\gamma}{25}\frac{\gamma+\gamma^{2,2}}{2\gamma-\alpha-\gamma^{2,2}}\,,\nonumber\\
\bar\beta_2&=&-\frac{9}{80}\tilde\gamma
+\frac{\tilde\gamma}{80}\frac{2\gamma+\gamma^{2,0}-1}{2\gamma-\alpha-(\gamma^{2,0}-1)/2}
-\frac{\tilde\gamma}{40}\frac{2\gamma+\gamma^{0,2}-1}{2\gamma-\alpha-(\gamma^{0,2}-1)/2}
-\frac{2\tilde\gamma}{25}\frac{\gamma+\gamma^{2,2}}{2\gamma-\alpha-\gamma^{2,2}}\,,\nonumber\\
\beta_3&=&-\frac{3}{80}\tilde\gamma
-\frac{\tilde\gamma}{160}\frac{2\gamma+\gamma^{2,0}-1}{2\gamma-\alpha-(\gamma^{2,0}-1)/2}
-\frac{\tilde\gamma}{160}\frac{2\gamma+\gamma^{0,2}-1}{2\gamma-\alpha-(\gamma^{0,2}-1)/2}
+\frac{11\tilde\gamma}{200}\frac{\gamma+\gamma^{2,2}}{2\gamma-\alpha-\gamma^{2,2}}\,, \nonumber\\
\beta_4&=&-\frac{3}{40}\tilde\gamma
-\frac{3\tilde\gamma}{160}\frac{2\gamma+\gamma^{2,0}-1}{2\gamma-\alpha-(\gamma^{2,0}-1)/2}
-\frac{3\tilde\gamma}{160}\frac{2\gamma+\gamma^{0,2}-1}{2\gamma-\alpha-(\gamma^{0,2}-1)/2}
+\frac{21\tilde\gamma}{200}\frac{\gamma+\gamma^{2,2}}{2\gamma-\alpha-\gamma^{2,2}}\,, \nonumber\\
\beta_5&=&\frac{3}{40}\tilde\gamma
+\frac{3\tilde\gamma}{160}\frac{2\gamma+\gamma^{2,0}-1}{2\gamma-\alpha-(\gamma^{2,0}-1)/2}
+\frac{3\tilde\gamma}{160}\frac{2\gamma+\gamma^{0,2}-1}{2\gamma-\alpha-(\gamma^{0,2}-1)/2}
-\frac{3\tilde\gamma}{200}\frac{\gamma+\gamma^{2,2}}{2\gamma-\alpha-\gamma^{2,2}}\,, \nonumber\\
\beta_6&=&\frac{3}{20}\tilde\gamma
+\frac{\tilde\gamma}{32}\frac{2\gamma+\gamma^{2,0}-1}{2\gamma-\alpha-(\gamma^{2,0}-1)/2}
-\frac{\tilde\gamma}{160}\frac{2\gamma+\gamma^{0,2}-1}{2\gamma-\alpha-(\gamma^{0,2}-1)/2}
-\frac{13\tilde\gamma}{200}\frac{\gamma+\gamma^{2,2}}{2\gamma-\alpha-\gamma^{2,2}}\,, \nonumber\\
\bar\beta_6&=&\frac{3}{20}\tilde\gamma
-\frac{\tilde\gamma}{160}\frac{2\gamma+\gamma^{2,0}-1}{2\gamma-\alpha-(\gamma^{2,0}-1)/2}
+\frac{\tilde\gamma}{32}\frac{2\gamma+\gamma^{0,2}-1}{2\gamma-\alpha-(\gamma^{0,2}-1)/2}
-\frac{13\tilde\gamma}{200}\frac{\gamma+\gamma^{2,2}}{2\gamma-\alpha-\gamma^{2,2}}\,, \nonumber\\
\beta_7&=&-\frac{3}{10}\tilde\gamma
-\frac{\tilde\gamma}{32}\frac{2\gamma+\gamma^{2,0}-1}{2\gamma-\alpha-(\gamma^{2,0}-1)/2}
-\frac{\tilde\gamma}{32}\frac{2\gamma+\gamma^{0,2}-1}{2\gamma-\alpha-(\gamma^{0,2}-1)/2}
-\frac{11\tilde\gamma}{200}\frac{\gamma+\gamma^{2,2}}{2\gamma-\alpha-\gamma^{2,2}}\,, \nonumber\\
\beta_8&=&-\frac{3}{20}\tilde\gamma
+\frac{\tilde\gamma}{160}\frac{2\gamma+\gamma^{2,0}-1}{2\gamma-\alpha-(\gamma^{2,0}-1)/2}
+\frac{\tilde\gamma}{160}\frac{2\gamma+\gamma^{0,2}-1}{2\gamma-\alpha-(\gamma^{0,2}-1)/2}
+\frac{19\tilde\gamma}{200}\frac{\gamma+\gamma^{2,2}}{2\gamma-\alpha-\gamma^{2,2}}\,,\nonumber\\
\beta_9&=&\frac{9}{40}\tilde\gamma
-\frac{\tilde\gamma}{40}\frac{2\gamma+\gamma^{2,0}-1}{2\gamma-\alpha-(\gamma^{2,0}-1)/2}
+\frac{\tilde\gamma}{80}\frac{2\gamma+\gamma^{0,2}-1}{2\gamma-\alpha-(\gamma^{0,2}-1)/2}
+\frac{\tilde\gamma}{25}\frac{\gamma+\gamma^{2,2}}{2\gamma-\alpha-\gamma^{2,2}}\,,\nonumber\\
\bar\beta_9&=&\frac{9}{40}\tilde\gamma
+\frac{\tilde\gamma}{80}\frac{2\gamma+\gamma^{2,0}-1}{2\gamma-\alpha-(\gamma^{2,0}-1)/2}
-\frac{\tilde\gamma}{40}\frac{2\gamma+\gamma^{0,2}-1}{2\gamma-\alpha-(\gamma^{0,2}-1)/2}
+\frac{\tilde\gamma}{25}\frac{\gamma+\gamma^{2,2}}{2\gamma-\alpha-\gamma^{2,2}}\,. \label{sphere41}
\end{eqnarray}
The first term proportional to $\tilde\gamma$ describes the influence of the order parameters 
themselves, while the other terms stand for the contributions of the center manifold. 
\subsection{Real Variables}

To investigate how the complex order parameter equations contribute to the
one-to-one retinotopy, we transform them to real variables according to
\begin{equation} \begin{array}{rclcrclcrcl}
\vspace{0.2cm}
u_0&=&U^{00}/\sqrt{2}&,&u_1&=&(U^{11}+U^{-1-1})/2&,&u_2&=&i(U^{11}-U^{-1-1})/2\\\vspace{0.2cm}
u_3&=&(U^{1-1}+U^{-11})/2&,&u_4&=&i(U^{1-1}-U^{-11})/2&,&u_5&=&(U^{01}-U^{0-1})/2\\
u_6&=&i(U^{01}+U^{0-1})/2&,&u_7&=&(U^{10}-U^{-10})/2&,&u_8&=&i(U^{10}+U^{-10})/2\,. 
\end{array}
\end{equation}
Then the equations of evolution for the real variables $u_i$ read
\begin{eqnarray}
\dot u_0&=&\Lambda u_0+2 \beta_1 u_0^3+2\bar\beta_2 u_0(u_5^2+u_6^2)+2\beta_2 u_0(u_7^2+u_8^2)
+2\beta_3 u_0(u_1^2+u_2^2+u_3^2+u_4^2)\nonumber\\
 & &+\sqrt{2}\,\beta_4(u_1u_5u_7+u_2u_5u_8+u_2u_6u_7+u_4u_6u_7-u_1u_6u_8-u_3u_5u_7-u_4u_5u_8
-u_3u_6u_8)\,,\label{kureellopgu0}\\
\dot u_1&=&\Lambda u_1+\sqrt{2}\,\beta_4 u_0(u_5u_7-u_6u_8)+\beta_5(u_3u_5^2-u_3u_6^2-2u_4u_5u_6)
-\beta_6u_1(u_5^2+u_6^2)-\bar\beta_6 u_1(u_7^2+u_8^2)\nonumber\\
 & &+2\beta_3u_0^2 u_1+\beta_5(u_3u_7^2-u_3u_8^2+2u_4u_7u_8)+\beta_7u_1(u_3^2+u_4^2)+\beta_8u_1(u_1^2+u_2^2)\,,\\
\dot u_2&=&\Lambda u_2+\sqrt{2}\,\beta_4 u_0(u_5u_8+u_6u_7)+\beta_5(u_5^2u_4-u_4u_6^2+2u_3u_5u_6)
-\beta_6u_2(u_5^2+u_6^2)-\bar\beta_6 u_2(u_7^2+u_8^2)\nonumber\\
 & &+2\beta_3u_0^2u_2-\beta_5(u_4u_7^2-u_4u_8^2-2u_3u_7u_8)+\beta_7u_2(u_3^2+u_4^2)+\beta_8u_2(u_1^2+u_2^2)\,,\\
\dot u_3&=&\Lambda u_3-\sqrt{2}\,\beta_4 u_0(u_5u_7+u_6u_8)+\beta_5(u_1u_5^2-u_1u_6^2+2u_2u_5u_6)
-\beta_6u_3(u_5^2+u_6^2)-\bar\beta_6 u_3(u_7^2+u_8^2)\nonumber\\
 & &+2\beta_3u_0^2u_3+\beta_5(u_1u_7^2-u_1u_8^2+2u_2u_7u_8)+\beta_7u_3(u_1^2+u_2^2)+\beta_8u_3(u_3^2+u_4^2)
\,,\label{kureellopgu3}\\
\dot u_4&=&\Lambda u_4+\sqrt{2}\,\beta_4 u_0(u_6u_7-u_5u_8)+\beta_5(u_2u_5^2-u_2u_6^2-2u_1u_5u_6)-\beta_6u_4(u_5^2
+u_6^2)-\bar\beta_6 u_4(u_7^2+u_8^2)\nonumber\\
 & &+2\beta_3u_0^2u_4+\beta_5(u_2u_8^2-u_2u_7^2+2u_1u_7u_8)+\beta_7u_4(u_1^2+u_2^2)+\beta_8u_4(u_3^2+u_4^2)\,,
\label{kureellopgu4}\\
\dot u_5&=&\Lambda u_5+2\bar\beta_2u_0^2u_5-\beta_9u_5(u_5^2+u_6^2)+2\beta_3 u_5(u_7^2+u_8^2)+\sqrt{2}
\,\beta_4u_0(u_1u_7-u_3u_7+u_2u_8-u_4u_8)\nonumber\\
 & &-\beta_6u_5(u_1^2+u_2^2+u_3^2+u_4^2)-2\beta_5(u_1u_4u_6-u_1u_3u_5-u_2u_4u_5-u_2u_3u_6)\,,\\
\dot u_6&=&\Lambda u_6+2\bar\beta_2u_0^2u_6-\beta_9u_6(u_5^2+u_6^2)+2\beta_3 u_6(u_7^2+u_8^2)+\sqrt{2}
\,\beta_4u_0(u_2u_7+u_4u_7-u_1u_8-u_3u_8)\nonumber\\
 & &-\beta_6u_6(u_1^2+u_2^2+u_3^2+u_4^2)-2\beta_5(u_1u_4u_5+u_1u_3u_6+u_2u_4u_6-u_2u_3u_5)\,,\\
\dot u_7&=&\Lambda u_7+2\beta_2u_0^2u_7-\bar\beta_9u_7(u_7^2+u_8^2)+2\beta_3 u_7(u_5^2+u_6^2)+\sqrt{2}
\,\beta_4u_0(u_1u_5-u_3u_5+u_2u_6+u_4u_6)\nonumber\\
 & &-\bar\beta_6 u_7(u_1^2+u_2^2+u_3^2+u_4^2)-2\beta_5(u_2u_4u_7-u_1u_3u_7-u_2u_3u_8-u_1u_4u_8)\,,\\
\dot u_8&=&\Lambda u_8+2\beta_2u_0^2u_8-\bar\beta_9u_8(u_7^2+u_8^2)+2\beta_3 u_8(u_5^2+u_6^2)-\sqrt{2}
\,\beta_4u_0(u_1u_6+u_3u_6-u_2u_5+u_4u_5)\nonumber\\
 & &-\bar\beta_6u_8(u_1^2+u_2^2+u_3^2+u_4^2)-2\beta_5(u_1u_3u_8-u_2u_3u_7-u_1u_4u_7-u_2u_4u_8)
\,.\label{kureellopgu8}
\end{eqnarray}
Note that the real order parameter equations (\ref{kureellopgu0})--(\ref{kureellopgu8}) 
follow according to
\begin{equation}
\dot u_i=-\frac{\partial V(\{u_j\})}{\partial u_i}
\end{equation}
from the potential
\begin{eqnarray}
V(\{u_j\})&=&-\frac{\Lambda}{2}\sum_{j=0}^8u_j^2-\frac{\beta_1}{2}u_0^4-\bar\beta_2 u_0^2(u_5^2+u_6^2)
-\beta_2 u_0^2(u_7^2+u_8^2)-\beta_3 u_0^2(u_1^2+u_2^2+u_3^2+u_4^2)\nonumber\\
 & &\hspace{-1.1cm}-\sqrt{2}\beta_4u_0(u_1u_5u_7+u_2u_5u_8+u_2u_6u_7+u_4u_6u_7-u_1u_6u_8
-u_3u_5u_7-u_4u_5u_8-u_3u_6u_8)\nonumber\\
 & &\hspace{-1.1cm}-\beta_5(u_5^2-u_6^2)(u_1u_3+u_2u_4)-\beta_5(u_7^2-u_8^2)(u_1u_3-u_2u_4)
-2\beta_5u_7u_8(u_1u_4+u_2u_3)\nonumber\\
 & &\hspace{-1.1cm}-2\beta_5u_5u_6(u_2u_3-u_1u_4)+\frac{1}{2}[\beta_6(u_5^2+u_6^2)
+\bar\beta_6(u_7^2+u_8^2)](u_1^2+u_2^2+u_3^2+u_4^2)-\frac{\beta_7}{2}(u_1^2+u_2^2)(u_3^2+u_4^2)\nonumber\\
 & &\hspace{-1.1cm}-\beta_3 (u_5^2+u_6^2)(u_7^2+u_8^2)-\frac{\beta_8}{4}\left[(u_1^2
+u_2^2)^2+(u_3^2+u_4^2)^2\right]+\frac{\beta_9}{4}(u_5^2+u_6^2)^2+\frac{\bar\beta_9}{4}(u_7^2+u_8^2)^2
\,. \label{kupot}
\end{eqnarray}
Naturally, a complete analytical determination of all stationary states of the real order 
parameter equations (\ref{kureellopgu0})--(\ref{kureellopgu8}) is
impossible. However, we are able to demonstrate that certain stationary states
admit for retinotopic modes. 
\subsection{Special Case}

To this end we consider the special case 
$u_1,u_2,u_5,u_6,u_7,u_8=0$. Then the equations (\ref{kureellopgu0}), (\ref{kureellopgu3}), 
and (\ref{kureellopgu4})
for the non-vanishing amplitudes $u_0$, $u_3$, $u_4$ reduce to 
\begin{eqnarray}
\dot u_0&=&\Lambda u_0+2 \beta_1 u_0^3+2\beta_3 (u_3^2+u_4^2)u_0\,,\nonumber\\
\dot u_3&=&\Lambda u_3+2\beta_3 u_0^2 u_3+\beta_8 (u_3^2+u_4^2)u_3\,,\nonumber\\
\dot u_4&=&\Lambda u_4+2\beta_3 u_0^2 u_4+\beta_8 (u_3^2+u_4^2)u_4\,.
\end{eqnarray}
Due to the relation
\begin{equation}
\frac{\dot u_3}{u_3}=\frac{\dot u_4}{u_4}
\end{equation}
one obtains constant phase-shift angles, i.e. it holds $u_3\propto u_4$. 
Therefore, the system of three coupled differential equations can be reduced to two variables. 
To this end we introduce the new variable 
\begin{equation} \label{kuneuvariab} 
\xi=\sqrt{u_3^2+u_4^2}\,,
\end{equation}
which leads to
\begin{eqnarray}
\dot u_0&=&\Lambda u_0+2 \beta_1 u_0^3+2\beta_3 \xi^2 u_0\,,\label{kuu0glg}\nonumber\\
\dot \xi&=&\Lambda \xi+2\beta_3 u_0^2 \xi +\beta_8 \xi^3\,.
\end{eqnarray}
The stationary solution, which corresponds to a coexistence of the two modes, 
is given by
\begin{equation} \label{kuu0xilsg}
u_0^2=-\frac{\Lambda}{2(\beta_3+\beta_8)}\,,\quad\xi^2=-\frac{\Lambda}{\beta_3+\beta_8}\,,
\end{equation}
where we used the relation $\beta_8=\beta_1+\beta_3$ following from (\ref{sphere41}). Demanding real 
amplitudes $u_0$, $\xi$ leads to the coexistence condition
\begin{equation} \label{kukoexungl1}
\beta_3+\beta_8<0\,.
\end{equation}
Furthermore, we require stability for this state. Therefore we consider
the corresponding potential $V(u_0,\xi)$, which can be read off from (\ref{kupot}) and 
(\ref{kuneuvariab}):
\begin{equation}
V(u_0,\xi)=-\frac{\Lambda}{2}(u_0^2+\xi^2)-\frac{\beta_1}{2}u_0^4-\beta_3 u_0^2\xi^2-\frac{\beta_4}{4}\xi^4\,.
\end{equation}
Stable states correspond to a minimum of $V$, which leads to the conditions 
\begin{equation} \label{kukoexungl2}
2\beta_3-\beta_8>0\,,\quad \beta_3-\beta_8>0\,.
\end{equation}
The inequalities (\ref{kukoexungl1}), (\ref{kukoexungl2}) can be summarized according to
\begin{equation}
\beta_8<0\,,\quad \beta_3<-\beta_8\,,\quad 2\beta_3>\beta_8\,.
\end{equation}
If they are valid, both the $u_0$- and the $\xi$-mode coexist. 
If we set $u_4=0$, without loss of generality, the solution reads in complex variables 
according to (\ref{kuneuvariab}) 
\begin{equation}
U^{00}=\sqrt{-\frac{\Lambda}{\beta_3+\beta_8}}\,,\qquad U^{1-1}=U^{-11}=-\sqrt{-\frac{\Lambda}{\beta_3+\beta_8}}\,.
\end{equation}
Thus, the unstable part (\ref{UEXP}) is given by
\begin{equation} \label{sphere66}
U(\hat t,\hat r)=\sqrt{-\frac{\Lambda}{\beta_3+\beta_8}}\,\Big[Y_{10}^T(\hat t\,) Y_{10}^R(\hat r)
-Y_{11}^T(\hat t\,)Y_{1-1}^R(\hat r)-Y_{1-1}^T(\hat t\,)Y_{11}^R(\hat r)\Big]\,.
\end{equation}
Using the Legendre addition theorem (\ref{kuaddtheorem}) reduces (\ref{sphere66}) to
\begin{equation} \label{kupl1}
U(\hat t,\hat r)=\sqrt{-\frac{\Lambda}{\beta_3+\beta_8}}\,P_1(\hat t\cdot\hat r)
\end{equation}
with $P_1(\hat t\cdot \hat r)=\hat t \cdot \hat r$. Thus, the unstable part is minimal, if 
$\hat t$ and $\hat r$ are antiparallel,
i.e. the distance of the corresponding points on the unit sphere is maximum.
Decreasing of the angle between $\hat t$ and $\hat r$ leads to increasing 
values of $U(\hat t,\hat r)$, and the maximum occurs for parallel unit vectors.
This justifies calling the mode (\ref{kupl1}) retinotopic.
\section{One-to-One Retinotopy} \label{11retino}
Now we investigate whether the generalized H{\"a}ussler equations (\ref{hslerkugel})
describe the emergence of a perfect one-to-one retinotopy between two spheres. 
To this end we follow the unpublished suggestions of Ref.~\cite{Malsburg4} and 
treat systematically the contribution
of higher modes. Because the Legendre functions form a complete orthogonal system 
(\ref{wagnerportho}), (\ref{wagnerpvollst}) for functions defined on the interval $[-1,+1]$, 
their products can always be written as linear combinations 
of Legendre functions. This motivates that the influence
of higher modes upon the connection weights, which obey
the generalized H{\"a}ussler equations (\ref{hslerkugel}), can be 
included by the ansatz
\begin{equation} \label{wagneransatz}
w(\sigma)=\sum_{l=0}^{\infty}(2l+1)Z_l P_l(\sigma)\,,
\end{equation}
where the amplitudes $Z_l$ are time dependent. 

\subsection{Recursion Relations}
Inserting (\ref{wagneransatz}) into the generalized H{\"a}ussler equations (\ref{hslerkugel}) 
and performing the integrals 
over the respective unit spheres leads to
\begin{equation}
\sum_{l=0}^{\infty}(2l+1)\dot Z_l P_l(\sigma)=\alpha\left[1-\sum_{l=0}^{\infty}(2l+1)Z_l P_l(\sigma)\right]
+\sum_{l=0}^{\infty}(2l+1)Z_l P_l(\sigma)\sum_{l'=0}^{\infty}(2l'+1)Z_{l'}f_{l'}^T f_{l'}^R [P_{l'}(\sigma)-Z_{l'}]
\,.\label{wagnerstern}
\end{equation}
The products of Legendre functions occuring in (\ref{wagnerstern}) can be 
reduced to linear combinations of single Legendre functions according to 
the standard decomposition \cite[8.915]{grad}
\begin{equation}
P_l(\sigma)P_{l'}(\sigma)=\sum_{k=0}^{l}A_{l,l',k}P_{l+l'-2k}(\sigma)\,,\quad l\leq l'
\end{equation}
with the coefficients
\begin{equation} \label{wagnerA}
A_{l,l',k}=
\frac{(2l'+2l-4k+1)\,a_{l'-k}a_k a_{l-k}}
{(2l'+2l-2k+1)\,a_{l+l'-k}}\,,
\quad a_k=\frac{(2k-1)!!}{k!}\,.
\end{equation}
Thus, contributions to the polynomial $P_{\tilde l}(\sigma)$ only occur iff the relation 
$k=(l+l'-\tilde l)/2$
is fulfilled. Furthermore, using the orthonormality relation (\ref{wagnerportho}) 
yields the following recursion relation for the amplitudes $Z_l$:
\begin{eqnarray}
(2l+1)\dot Z_l&=&\alpha[\delta_{l,0}-(2l+1)Z_l]-(2l+1)Z_l(Z_0^2+3f_1^T f_1^R Z_1^2)
+\sum_{l'=0}^{\infty}(2l'+1)Z_{l'}\left[\sum_{l''=0}^l (2l''+1)Z_{l''}f_{l''}^T f_{l''}^R \right.\nonumber\\
 & &\left.\times\sum_{k=0}^{l''}A_{l',l'',k}\delta_{k,(l'+l''-l)/2}+\sum_{l''=l'+1}^{\infty}(2l''+1)
Z_{l''}f_{l''}^T f_{l''}^R \sum_{k=0}^{l'}A_{l',l'',k}\delta_{k,(l'+l''-l)/2} \right]\,. \label{128}
\end{eqnarray}
Note that Eq.~(\ref{128}) cannot be solved analytically for arbitrary expansion coefficients 
$f_l^T$, $f_l^R$ 
of the cooperativity functions. Therefore, we restrict ourselves from now on to
a special case.

\subsection{Special Cooperativity Functions}
For simplicity we assume that the 
expansion of the cooperativity functions (\ref{kuct}) breaks down after the first order:
\begin{equation} \label{129}
c_T(\hat t\cdot\hat t')=\frac{1}{4\pi}[1+3 f_1^T P_1(\hat t\cdot\hat t')]\,,\qquad c_R(\hat r\cdot\hat r')
=\frac{1}{4\pi}[1+3 f_1^R P_1(\hat r\cdot\hat r')]\,.
\end{equation} 
With this choice the recursion relation (\ref{128}) for $l=0$ reduces to
\begin{equation} \label{wagnerz0}
\dot Z_0=-(\alpha+Z_0^2+3\gamma Z_1^2)(Z_0-1)\,,
\end{equation}
where we have used again the abbreviation $\gamma=f_1^T f_1^R$.
For $l\not=0$, by taking into account (\ref{wagnerA}), we obtain 
\begin{equation} \label{wagner3stern}
\dot Z_l=-(\alpha+Z_0^2+3\gamma Z_1^2)Z_l+Z_0 Z_l+3\gamma Z_1\,\frac{lZ_{l-1}+(l+1)Z_{l+1}}{2l+1}\,.
\end{equation}
The long-time behavior of the system corresponds to its stationary states. 
They are determined by $Z_0=1$ from (\ref{wagnerz0}), whereas
(\ref{wagner3stern}) leads to a nonlinear recursion relation for the amplitudes 
$Z_l$ with $l\not=0$. However, by introducing the variable
\begin{equation} \label{wagnerualpha}
u=\frac{\alpha+3\gamma Z_1(u)^2}{3\gamma Z_1(u)}\,,
\end{equation}
this nonlinear recursion relation can be formally transformed into the linear one
\begin{equation} \label{wagnerrek}
(l+1)Z_{l+1}(u)=(2l+1)uZ_l(u)+lZ_{l-1}(u)\,,\quad l\geq 1\,.
\end{equation}
Thus, solving the nonlinear recursion relation (\ref{wagner3stern}) amounts to
solving the linear recursion relation (\ref{wagnerrek}) for $Z_l(u)$ in such
a way that the self-consistency condition (\ref{wagnerualpha}) is fulfilled.

\subsection{Generating Function}
To determine the amplitudes $Z_l(u)$ we calculate their
generating function
\begin{equation} \label{wagnererz}
E(x,u)=\sum_{l=0}^{\infty}Z_l(u)x^l\,,
\end{equation}
where we have the normalization 
\begin{equation} \label{wagnerex0}
E(0,u)=Z_0(u)=1\,.
\end{equation}
Multiplying both sides of (\ref{wagnerrek}) with $x^l$ and summing over $l\geq 1$
leads to an inhomogeneous nonlinear partial differential equation of first order for 
the generating function: 
\begin{equation} \label{wagnerinh}
(x^2-2ux+1)\,\frac{\partial E(x,u)}{\partial x}=(u-x)E(x,u)+Z_1(u)-u\,.
\end{equation}
At first, we consider the homogeneous equation corresponding to (\ref{wagnerinh}):
\begin{equation}
(x^2-2ux+1)\,\frac{\partial E_{\rm hom}(x,u)}{\partial x}=(u-x)E_{\rm hom}(x,u)\,.
\end{equation}
It is solved by the method of separating variables, yielding
\begin{equation}
E_{\rm hom}(x,u)=\frac{K(u)}{\sqrt{x^2-2ux+1}}\,,
\end{equation}
where $K(u)$ is an integration constant. 
Afterwards, we determine a particular solution of the inhomogeneous equation 
(\ref{wagnerinh}) by using the method of varying constants. Using the ansatz
\begin{equation}
E_{\rm part}(x,u)=\frac{K(x,u)}{\sqrt{x^2-2ux+1}}
\end{equation}
leads to the differential equation
\begin{equation}
\frac{\partial K(x,u)}{\partial x}=\frac{Z_1(u)-u}{\sqrt{x^2-2ux+1}}\,,
\end{equation}
which is solved by using \cite[2.261]{grad}:
\begin{equation}
K(x,u)=\frac{[Z_1(u)-u]\ln[2\sqrt{x^2-2ux+1}+2(x-u)]}{\sqrt{x^2-2ux+1}}\,.
\end{equation}
Thus, the complete solution $E(x,u)=E_{\rm hom}(x,u)+E_{\rm part}(x,u)$ of
Eq.~(\ref{wagnerinh}) reads as follows:
\begin{equation}
E(x,u)=\frac{K(u)+[Z_1(u)-u]\ln[2\sqrt{x^2-2ux+1}+2(x-u)]}{\sqrt{x^2-2ux+1}}\,.
\end{equation}
Furthermore, using the normalization condition (\ref{wagnerex0}) fixes the integration 
constant to $K(u)=1-[Z_1(u)-u]\ln(2-2u)$.
Thus, the generating function is finally given by
\begin{equation} \label{wagnererzeugende}
E(x,u)=\frac{1+[Z_1(u)-u]\ln\displaystyle{\frac{\sqrt{x^2-2ux+1}+x-u}{1-u}}}{\sqrt{x^2-2ux+1}}\,.
\end{equation}

\subsection{Decomposition}
We now determine the unknown amplitudes $Z_l(u)$.
From the mathematical literature
it is well-known that the recursion relation (\ref{wagnerrek}) holds both for the
Legendre functions of first kind $P_l(u)$ and second kind $Q_l(u)$, respectively \cite{grad}. 
Thus, we expect that the generating function (\ref{wagnererzeugende}) can be
represented as a linear combination of the generating functions of the Legendre
functions of both first and second kind, 
which are given by \cite[8.921]{grad} and \cite[8.791.2]{grad}:
\begin{eqnarray}
E_P(x,u)&=&\sum_{l=0}^{\infty}P_l(u)x^l=\frac{1}{\sqrt{x^2-2ux+1}}\,,\label{Pgen}\\
E_Q(x,u)&=&\sum_{l=0}^{\infty}Q_l(u)x^l=
\frac{\ln\displaystyle{\frac{\sqrt{x^2-2ux+1}+u-x}{\sqrt{u^2-1}}}}{\sqrt{x^2-2ux+1}}
\,.\label{Qgen}
\end{eqnarray}
Indeed, taking into account the explicit form of the Legendre function of second kind 
for $l=0$~\cite{arf}
\begin{equation} \label{wagnerq0}
Q_0(u)=\frac{1}{2}\ln \frac{u+1}{u-1}\,,
\end{equation}
the generating function (\ref{wagnererzeugende}) decomposes according to
\begin{equation}
E(x,u)=\{1+[Z_1(u)-u]Q_0(u)\}E_P(x,u)-[Z_1(u)-u]E_Q(x,u)\,.
\end{equation}
Inserting (\ref{Pgen}), (\ref{Qgen}) and performing a comparison with (\ref{wagnererz}) 
then yields the result
\begin{equation} \label{wagnerzl}
Z_l(u)=\{1+[Z_1(u)-u]Q_0(u)\}P_l(u)-[Z_1(u)-u]Q_l(u)\,.
\end{equation}
Thus, the amplitudes $Z_l(u)$ turn out to be linear combinations of $P_l(u)$ and $Q_l(u)$. 
To fix the yet undetermined amplitude $Z_1(u)$ in the expansion coefficients of
(\ref{wagnerzl}), we have to take into account the boundary condition that
the sum in the ansatz (\ref{wagneransatz}) has to converge.

\subsection{Boundary Condition}
\begin{figure}[t!]
 \centerline{\includegraphics[scale=0.9]{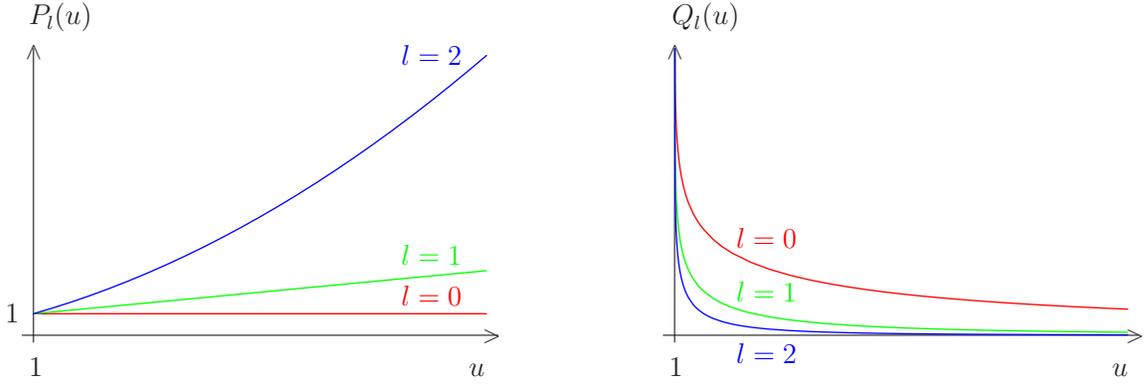}}
 \caption[Legendre-Polynome]{\label{pqlegendre} \small The Legendre functions of first and 
second kind $P_l(u)$ and $Q_l(u)$ for $u>1$. 
We have $P_l(1)=1$, whereas $Q_l(u)$ diverges for $u\downarrow 1$. 
Important for the boundary condition of $Z_l(u)$ is the different behavior
for increasing values of $l$: $P_l(u)$ diverges according to (\ref{wagnerpldiv}), whereas 
$Q_l(u)$ converges to zero.}
\end{figure}
Because the Legendre functions $P_l(\sigma)$ do not vanish with increasing $l$, 
we must require 
\begin{equation}
\lim_{l\to \infty}Z_l(u)=0\,.
\end{equation}
The series of Legendre functions of first kind $P_l(u)$ with fixed $u>1$ diverges for 
$l\to \infty$ according to \cite[8.917]{grad}
\begin{equation} \label{wagnerpldiv}
P_0(u)<P_1(u)<P_2(u)<\ldots<P_n(u)<\ldots\,,\quad u>1\,.
\end{equation}
The Legendre functions of second kind $Q_l(u)$, however, converge to zero 
(see Figure~\ref{pqlegendre}). Thus, performing the limit
$l\to \infty$ in Eq.~(\ref{wagnerzl}), we obtain
\begin{equation}
1+[Z_1(u)-u]Q_0(u)=0\,.
\end{equation}
From the explicit form \cite{arf} $Q_1(u)=uQ_0(u)-1$
it follows that $Z_1(u)$ is fixed according to
\begin{equation} \label{wagnerz1}
Z_1(u)=\frac{Q_1(u)}{Q_0(u)}\,.
\end{equation}
With this we obtain that 
the result (\ref{wagnerzl}) finally reads
\begin{equation} \label{wagnerzlq}
Z_l(u)=\frac{Q_l(u)}{Q_0(u)}\,,
\end{equation}
which is not valid only for $l\not=0$ but also for $l=0$ due to (\ref{wagnerex0}).

\subsection{Connection Weight}
Inserting (\ref{wagnerzlq}) into (\ref{wagneransatz}) yields the following 
solution for the connection weight:
\begin{equation}
w(\sigma)=\frac{1}{Q_0(u)}\sum_{l=0}^{\infty}(2l+1)Q_l(u)P_l(\sigma)\,.
\end{equation}
Using the identity \cite[8.791.1]{grad} 
\begin{equation}
\sum_{l=0}^{\infty}(2l+1)Q_l(u)P_l(\sigma)=\frac{1}{u-\sigma}
\end{equation}
and (\ref{wagnerq0}), we obtain for the connection weight 
\begin{equation} \label{117}
w(\sigma)=\frac{2}{u-\sigma}\left(\ln\frac{u+1}{u-1}\right)^{-1}\,.
\end{equation}
Note that integrating (\ref{117}) over the unit sphere leads to 
\begin{equation} 
\int\limits_{0}^{2\pi}\!\!d\varphi\int\limits_{-1}^{+1}\!\!d\sigma\,w(\sigma)=4\pi\,,
\end{equation}
i.e. the total connection weight coincides with the measure (\ref{kumass}). \\

On the other hand we have to take into account that the self-consistency 
condition (\ref{wagnerualpha}) yields an explicit relation between the variable $u$ 
and the control parameter $\alpha$. 
Indeed, we infer from (\ref{wagnerualpha}) 
and (\ref{wagnerz1}) the following transcendental relation between
$\alpha$ and $u$
\begin{equation} \label{wagnera2g}
\frac{\alpha}{\gamma}=-\frac{2}{3}\left(\ln\frac{u+1}{u-1}\right)^{-1}
\left[2\left(\ln\frac{u+1}{u-1}\right)^{-1}-u\right]\,, 
\end{equation}
which is depicted in Figure~\ref{alphavonu}a. 

\subsection{Limiting Cases}
\begin{figure}[t!]
 \centerline{a) \includegraphics[scale=0.7]{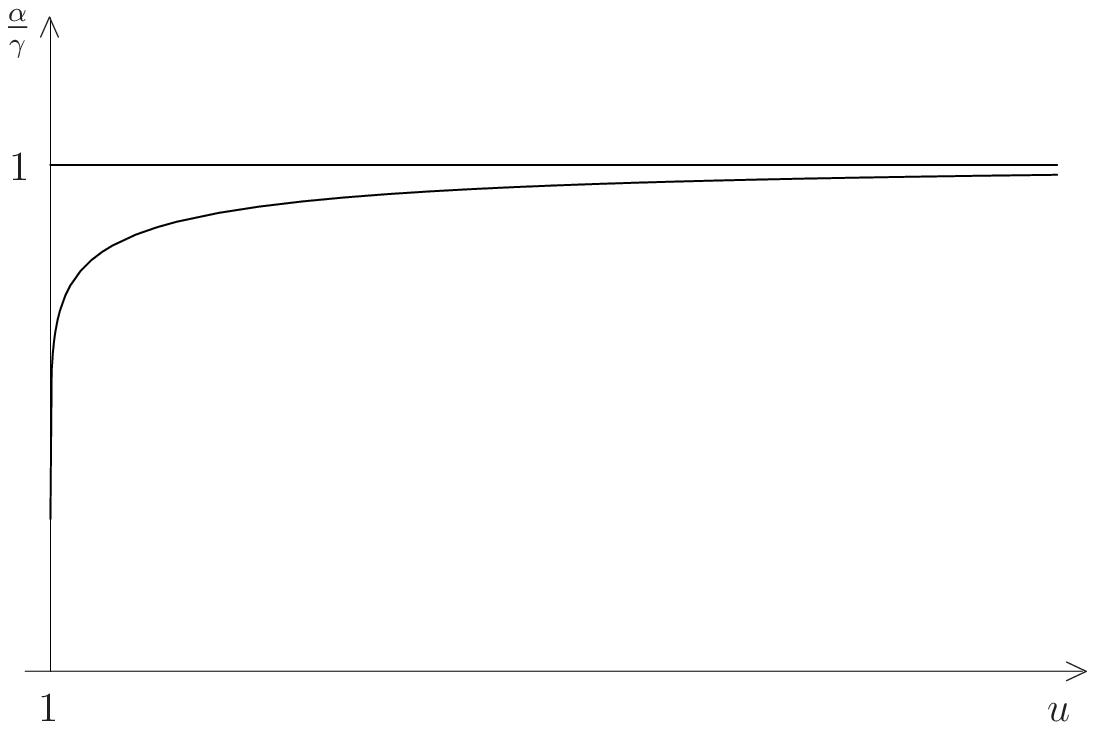} \hspace{0.5cm} b) 
\includegraphics[scale=0.7]{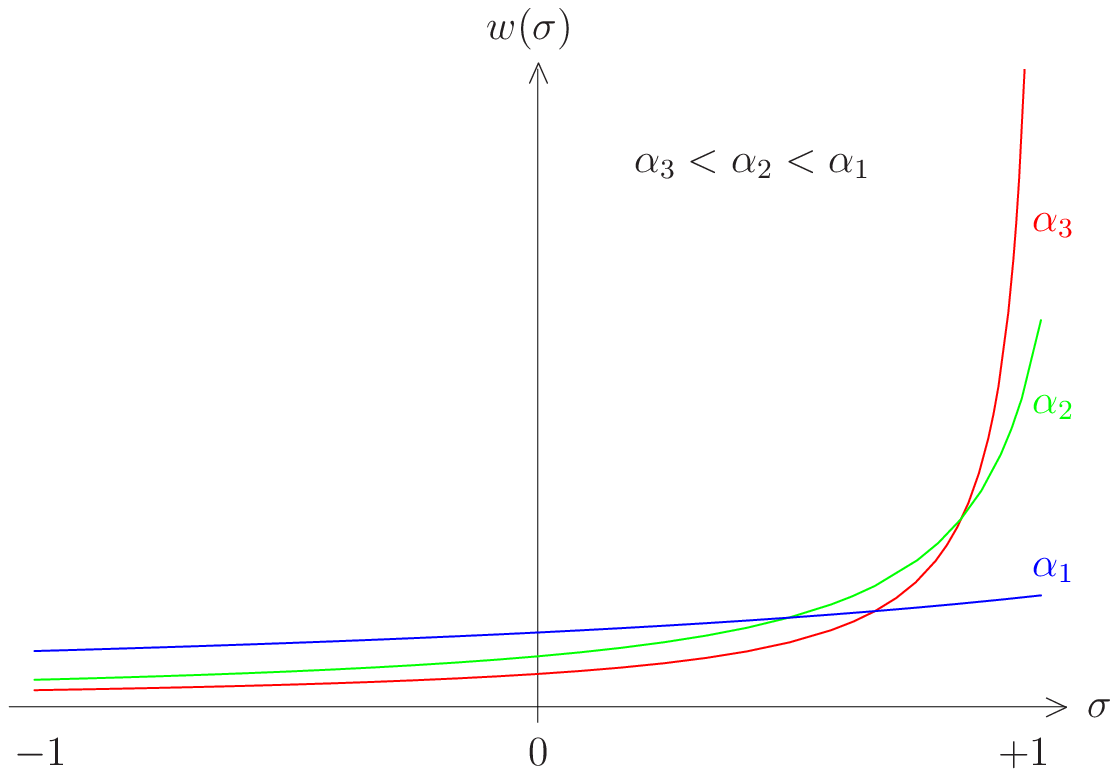}}
 \caption[Kontrollparameter]{\label{alphavonu} \small a) Relation (\ref{wagnera2g}) 
between the control parameter $\alpha$ and the variable $u$. b) The 
connection weight for different values of the control parameter $\alpha$. For decreasing 
values of $\alpha$ the connection weight around $\sigma=+1$ is growing. In the limiting case 
$\alpha\to 0$ the connection weight $w(\sigma)$
becomes Dirac's delta function (\ref{kudeltafkt}).}
\end{figure}
The limiting value of (\ref{wagnera2g}) for $u\to \infty$ is determined with the help of
the expansion \cite[1.513]{grad}
\begin{equation}
\ln\frac{1+x}{1-x}=2\sum_{k=1}^{\infty}\frac{1}{2k-1}\,x^{2k-1}\,,\quad x^2<1\,,
\end{equation}
and reads
\begin{equation} \label{uinft}
\lim_{u\to\,\infty}\alpha=\gamma\,.
\end{equation}
Thus, we conclude that the case $u\to\,\infty$  
corresponds to the instability point $\alpha_c=f_1^T f_1^R$, 
which was obtained from the linear stability analysis in 
Section~\ref{linanalys}. Correspondingly, using again (\ref{wagnera2g}), we observe
that the connection weight (\ref{117}) coincides in the limit $u\to \infty$ with
a uniform distribution: 
\begin{equation}
\lim_{\alpha\uparrow \alpha_c} w(\sigma)=1\,.
\end{equation}
Another biological important special case is $u\downarrow 1$, where we obtain
from (\ref{wagnera2g}) 
\begin{equation} \label{u1}
\lim_{u\downarrow 1}\alpha=0\,.
\end{equation}
Furthermore, considering the limit $u\downarrow 1$ in (\ref{117}) for $\sigma\not=u$, we obtain
\begin{equation}
\lim_{u\downarrow 1}\frac{2}{u-\sigma}\left(\ln\frac{u+1}{u-1}\right)^{-1}=0\,.
\end{equation}
On the other hand, integrating (\ref{117}) for $u\downarrow 1$ over $\sigma$ yields
\begin{equation}
\lim_{u\downarrow 1}\int\limits_{-1}^1 \frac{2}{u-\sigma}\left(\ln\frac{u+1}{u-1}\right)^{-1} d\sigma=2\,.
\end{equation}
Therefore, we conclude that the connection weight $(\ref{117})$ becomes 
in this limit Dirac's delta function:
\begin{equation} \label{kudeltafkt}
\lim_{\alpha\downarrow 0} w(\sigma)=4\delta(\sigma-1)\,.
\end{equation}
Thus, decreasing the control parameter $\alpha$ means that the projection
between two spheres becomes sharper and sharper (see Figure~\ref{alphavonu}b).
A perfect one-to-one retinotopy is achieved for $\alpha=0$ when the uniform and
undifferentiated formation of new synapses onto the tectum is completely terminated.

\section{Summary}
In this series of three papers we have analyzed in detail the
self-organized formation of retinotopic projections between
manifolds of different geometries. Applying our generalized
H{\"a}ussler equations \cite{gpw1} to Euclidean manifolds \cite{gpw2}, and to
spheres in the present paper, led to remarkably analogous results.
Both for one-dimensional strings and for spheres we have
furnished proof that our generalized H{\"a}ussler equations describe,
indeed, the emergence of a perfect one-to-one retinotopy. Furthermore,
we have shown in both cases that the underlying order parameter
equations follow from a potential dynamics and do not contain
quadratic terms. However, in contrast to strings, spherical manifolds
represent a more adequate description for retina and tectum.
Therefore, the present paper represents an essential progress in
the understanding of the ontogenetic development of neural connections
between retina and tectum.

\end{document}